\documentstyle{article}
\title{\LARGE Kepler problem  in deformed (quantum) four-dimensional space in non relativistic limit with Galilei group of motion}
\author{A.~N.~Leznov\thanks{ Universidad Autonoma del Estado de Morelos, CCICAp,Cuernavaca, Mexico}} \date{}

\newcommand{\rig}[2]{\stackrel{#2\rightarrow}{#1}}

\begin{document}
\maketitle

\maketitle
\title{\LARGE Kepler problem  in deformed (quantum) four-dimensional space in non relativistic limit with Galilei group of motion}
\begin{abstract}
It is shown that Kepler problem in  deformed (quantum) four-dimensional space in non relativistic limit is integrable in quadratures. In non relativistic limit group of motion of quantum space coincide with Galilei one.
\end{abstract}

\section{Introduction}

In previous paper we have noticed that in the space of constant curvature in three dimension equations of motion considered in numerous papers after the pioneer paper of E.Schrodinger \cite{SCH},\cite{IS},\cite{HIG},\cite{GKO} are not Galiley invariant. And thus results of these papers may be considered as very nice exercise for a students. It is very interesting that it is not possible the spaces of constant curvature in three dimension wide in such way to include the Galiley group of motion. As it follows from calculations of \cite{0} the three dimensional limit from Lorenz invariant quantum space leads to Snyder theory \cite{I} with non commuting coordinates but not impulses. 

In the present paper we would like investigate Kepler problem in deformed (quantum) space with equations invariant to Galiley group of motions -- one of most important laws of non relativistic dynamics. 

\section{Quantum spaces in non relativistic domain}

The non relativistic limit of Lorenz invariant quantum space was found and presented in \cite{0}. To obtain this formulae it is necessary in formulae of Lorenz invariant quantum space time \cite{LEZM} to do substitution  $x_4\to ct, p_4\to mc+{\epsilon\over c}, \rig{f}{}\to c\rig{f}{}, M^2\to M^2, L^2\to c^2T^2, H\to c^2 \nu$, and take the limit $c\to \infty$,
where $c$ is the velocity of the light $M^2,T^2,\nu$ some constants of non relativistic theory with the dimension of $(impulse)^2$, $(time)^2$ and mass-second correspondingly, $m$ c number constant with dimension of mass. Result is the following algebra of the 16  observable (we include among them $m$--the mass of the particle, which can considered as a center element of the algebra)
$$
[\epsilon,t]=ihI,\quad [\epsilon,\rig{p}{}]=ih{\rig{f}{}\over T^2},\quad [\epsilon,\rig{x}{}]=ih{\rig{f}{}\over \nu},\quad [\epsilon,I]=ih({t\over T^2}-{m\over \nu}),
$$
$$
[\epsilon,\rig{f}{}]=ih\rig{p}{}[\epsilon,\rig{l}{}]=0,\quad [\epsilon,\rig{f}{}]=ih\rig{p}{}[t,\rig{p}{}]=0,\quad [t,\rig{x}{}]=ih{\rig{f}{}\over M^2},\quad [t,I]=ih{m\over M^2},
$$
$$
[I,\rig{p}{}]=[I,\rig{l}{}]=[I,\rig{f}{}]=0,\quad [I,\rig{x}{}]=ih{\rig{p}{}\over M^2},\quad
[p_{\alpha},p_{\beta}]=0,\quad [x_{\alpha},x_{\beta}]=ih{\epsilon_{\alpha,\beta,\gamma}l_{\gamma\over M^2}}
$$
\begin{equation}
[t,\rig{f}{}]=[t,\rig{l}{}]=0,\quad [p_{\alpha},x_{\beta}]=ih\delta_{\alpha,\beta}I,\quad [f_{\alpha},x_{\beta}]=ih\delta_{\alpha,\beta}t,\quad [f_{\alpha},p_{\beta}]=ih\delta_{\alpha,\beta}m \label{NR}
\end{equation}
\begin{equation}
1=I^2-{t^2\over T^2}+{2mt\over \nu}+{1\over M^2}(-(\rig{p}{})^2+{(\rig{f}{})^2\over T^2}+2m\epsilon)\label{DC}
\end{equation}
$$
I\rig{f}{}=t \rig{p}{}+ \rig{x}{}m,\quad I\rig{l}{}=[\rig{x}{}\times \rig{p}{}]
$$
(\ref{NR}) are commutation relations of quantum space by itself, (\ref{DC}) additional conditions which are self consistent with algebra (\ref{NR}) and lead to usual Mincovski space-time with Poincare group of motion in the limit of infinite dimensional parameters of the quantum space. Below commutatores without $ih$ in the right hand side must be understandable as Poisson brackets.
  
As reader can see that sub algebra of (\ref{NR}) for 10 variables $\rig{p}{},\rig{x},\rig{l}{},I$  coincides with considered in \cite{LM} under condition $L^2\to \infty,S\to \infty$. And thus the case of the space of constant curvature considered by Schrodinger \cite{SCH} can't be deformed to Galilei invariant algebra. 
 
\section{Preliminary calculation and notation}

In the present paper we would like to consider Kepler problem of classical dynamics with
Poisson brackets for dynamical variables (\ref{NR}) and obvious exchanging 
${1\over ih} [A,B]\to \{A,B\}$ (compare with solution of the same problem for algebra of the constant curvature in \cite{LM}). 

If we want have some dynamical system with conserved energy it is necessary at first find among dynamical variables those, which commute (have zero Poisson brackets) with the energy $\epsilon$ from (\ref{NR}). It is easy to check that quadratical combination $ (\rig{p}{})^2-{(\rig{f}{})^2\over T^2}$ is commutative with energy $\epsilon$ and both vectors  $\rig{p}{}, \rig{f}{}$ and thous the hamiltonian of free motion can be chosen in a form \cite{0}
$$
H=2m\epsilon-(\rig{p}{})^2+({\rig{f}{})^2\over T^2}  
$$
where $m$ is the mass of the moving object.
The generalization of coordinates commuting with $\epsilon$ possible only one vector combination
$$
\rig{{\tilde x}}{}=\rig{x}{}-{T^2\over \nu}\rig{p}{}
$$
Thus if we choose Hamiltonian function in a form
$$
H=2m\epsilon-(\rig{p}{})^2+{(\rig{f}{})^2\over T^2}+U(\rig{{\tilde x}}{})
$$
we will have theory invariant to the transformation of Galilei group of deformed space with conserved energy.
From (\ref{NR}) it follows that tilde coordinates are not commuted but
$$
[\tilde x_{\alpha},\tilde x_{\beta}]=ih\epsilon_{\alpha,\beta,\gamma}{l_{\gamma}\over M^2},
\quad [p_{\alpha},\tilde x_{\beta}]=ih\delta_{\alpha,\beta}I
$$ 
If we would like to consider spherically symmetrical problem Hamiltonian function must depend 
only on $ \sum (\tilde x_{\beta})^2$ and if we assume that potential interaction part $U$ commute with all components vector $\rig{{x}}{}$ then potential function must be chosen in a form
$(\rig{{\tilde x}}{})^2+{(\rig{l}{})^2\over M^2}$. The last expression is the Kazimir operator of four dimensional rotating group with 6 generators $\rig{{\tilde x}}{},(\rig{l}{}$.

Finally we in what follows we will investigate the Kepler problem with Hamiltonian function
\begin{equation}
H=2m\epsilon-(\rig{p}{})^2+{(\rig{f}{})^2\over T^2}+{2m\alpha\over \sqrt {(\rig{{\tilde x}}{})^2+{(\rig{l}{})^2\over M^2}}}\label{HAM}
\end{equation}

\section{Algebra of spherically invariant variable}

All spherically invariant variables may be united into elements of two mutually commutative algebras of $O(1.2)$
groups in a following way ( $p^2\equiv (\rig{p}{})^2$, $l=\sqrt {(\rig{l}{})^2}$ and so on , below $t\equiv t-{mT^2\over \nu}$)
$$
T{p^2\over 2m}=s_+^1={l\over 2} \cosh \tau e^{\rho},\quad {f^2\over 2mT}=s_-^1={l\over 2} \cosh \tau e^{-\rho}, \quad {pf\over m}=s_0^1=l \sinh \tau
$$
\begin{equation}
-T{M^2I^2\over 2m}=s_+^2={l\over 2}R e^a,\quad -{M^2t^2\over 2Tm}=s_-^2={l\over 2}R e^{-a}, \quad-{M^2It\over m}=s_0^2=l R \label{PR}
\end{equation}
where $\tau,\rho,R,a$ are arbitrary no dimensional parameters.

Using (\ref{NR}) it is not difficult to check that two vectors $s^1,s^2$ are mutually commutative and commutation relations between its components are usual relation of algebra $O(1.2)$. Namely 
\begin{equation}
[s_+^i,s_-^i]=s_0^i ,\quad [s_0^i,s_{\pm}^i]=\mp2s_{\pm}^i,\quad s_+^1s_-^1-{1\over 4}(s_0^1)^2=
{1\over 4}l^2, \quad  s_+^2s_-^2-{1\over 4}(s_0^2)^2=0 \label{CR}
\end{equation}
The last relation in a formula above are corresponding Kazimir operators for representation of $O(1.2)$ algebras.

Let us do some manipulations with the expression under the root in Hamiltonian function (\ref{HAM}). 
we have in a consequence, using (\ref{DC}) we exchange $\rig{x}{}={t\over m}\rig{p}{}-
{I\over m}\rig{f}{}$,$\rig{l}{}={1\over m}[\rig{p}{}\times \rig{f}{}]$
$$
r^2\equiv (\rig{{\tilde x}}{})^2+{(\rig{l}{})^2\over M^2}={p^2t^2-2(pf)It+f^2I^2\over m^2}+{p^2f^2-(pf)^2\over M^2m^2}
$$ 
$$
={4\over M^2}[({p^2+M^2I^2\over 2m})({f^2+M^2t^2\over 2m})-{1\over 4}{(M^2It+pf)^2\over m^2}]=
$$
$$
{4\over M^2}[(s_+^1-s_+^2)(s_-^1-s_-^2)-{1\over 4}(s_0^1-s_0^2)^2]=
$$
or
\begin{equation}
r^2={4\over M^2}[(s_+^1s_-^1-{1\over 4}(s_0^1)^2)-((s_+^1s_-^2+s_-^1s_+^2-{1\over 2}s_0^1s_0^2)]
\label{R}
\end{equation} 
Now we are ready to present the Hamiltonian (\ref{HAM}) of a problem in terms of ''generators'' of $O(1.2)$ algebra in a following way
\begin{equation}
H=2m\epsilon-{2m\over T}(s_+^1-s_-^1)+{2m\alpha\over \sqrt {{4\over M^2}[(s_+^1s_-^1-{1\over 4}
(s_0^1)^2)-((s_+^1s_-^2+s_-^1s_+^2-{1\over 2}s_0^1s_0^2)]}}\label{HAM1}
\end{equation}
Using (\ref{DC}) Hamiltonian function (\ref{HAM1}) may be rewritten in equivalent form
\begin{equation}
H=M^2(1-{m^2T^2\over \nu^2})+{2m\over T}(s_+^2-s_-^2)+{m\alpha\over \sqrt {{4\over M^2}[(s_+^1s_-^1-{1\over 4}(s_0^1)^2)-((s_+^1s_-^2+s_-^1s_+^2-{1\over 2}s_0^1s_0^2)]}}\label{HAM11}
\end{equation}
From the last expression it is obvious that Poisson brackets of the first term of last formulae
with all components of the vector $\rig{s^1}{}$ equal to zero and the second term is commutative with all components of the vector $\rig{s^1+s^2}{}$. This fact essentially simplified calculations in the next section.

\section{Equations of motion and its resolving}

Differentiation on proper time will be denoted by dot index under corresponding variable. From  
(\ref{HAM1}),(\ref{HAM11}) and (\ref{CR}) it follows the system of equations for unknown components of
 $s^1,s^2$ (we include factor $\sqrt {{4\over M^2}}$ into $\alpha$). Thirst of all $\dot \epsilon=
[H,\epsilon]=0$ and thus energy is the integral of the motion. Further 
\begin{equation}
\dot {(s_+^1+s_+^2)}={2m\over T}s_0^2,\quad \dot {(s_-^1+s_-^2)}={2m\over T}s_0^2,\quad 
\dot {(s_0^1+s_0^2)}={4m\over T}(s_-^2+s_+^2) \label{FT}
\end{equation}
\begin{equation}
\dot {(s_+^1)}=-{m\alpha(s_0^1s_+^2-s_+^1s_0^2)\over 4r^3M^2},\quad \dot {(s_-^1)}={m\alpha(s_0^1s_-^2-s^1_-s_0^2)\over 4r^3M^2},\quad \dot {(s_0^1)}={m\alpha(s_+^1s^2_-s_+^2s_+^1)\over 2r^3M^2} 
\label{ST}
\end{equation}

First two equations of (\ref{FT}) lead to integral $-s_+^2+s_-^2-s_+^1+s_-^1$ which coincides with first additional condition of (\ref{DC}) and thus we have $-s_+^2+s_-^2-s_+^1+s_-^1={M^2T\over 2m}(1-({mT\over \nu})^2-{2\epsilon m\over M^2})\equiv \Theta.$ Now using (\ref{R}) we calculate derivative of $r^2$.  We have using (\ref{R})
\begin{equation}
\dot r^2=[H,r^2]={2m\over T}{4\over M^2}[(s_+^2-s_-^2),r^2]={2m\over T}{4\over M^2}
[-(s_+^2+s_-^2)s_0^1+(s_+^1+s_-^1)s_0^2]\label{DR}
\end{equation}
After reducing of two first equations from (\ref{ST}) we obtain second integral of motion
\begin{equation}
E_1=2m\epsilon-{m^2\over T}(s_+^1-s_-^1)+{m\alpha\over r}\label{E1}
\end{equation}
In combination with the first integral obtained above we represent it in equivalent form
\begin{equation}
E_1=M^2(1-{m^2T^2\over \nu^2})+{2m\over T}(s_+^2-s_-^2)+{m\alpha\over r}\label{E2}
\end{equation}

Now let us resolve two remaining equations of (\ref{FT}). After summation two first equations we come to system
$$
\dot {(s_+^1+s_+^2+s_-^1+s_-^2)}={4m\over T}s_0^2,\quad 
\dot {(s_0^1+s_0^2)}={4m\over T}(s_-^2+s_+^2) 
$$
Using parameterization  (\ref{PR}) for second vector $s^2$ we rewrite the system above
$$
\dot {(s_+^1+s_-^1)}+\cosh a\dot {lR} +\dot a \sinh a lR={4m\over T}lR,\quad 
\dot {(s_0^1)}+\dot {lR}={4m\over T}\cosh a lR
$$
After some trivial manipulations we obtain (${T\over 2ml}(E_1-M^2(1-{m^2T^2\over \nu^2})-{m\alpha\over r})=R\sinh a$ from (\ref{E2}) ). 
\begin{equation}
-{m^2\alpha \over 2M^2T r^3}{r^2-{l^2\over M^2}\over 1-{m^2T^2\over \nu^2}+{{m\alpha\over r}-E_1\over M^2}}-{\dot a\over \sinh a}={4m\over T}
\label{TR}
\end{equation}
To obtain equation for $r$ let us write dawn following three equations (\ref{R}),(\ref{E1}),(\ref{E2}), (\ref{DR}) in which we parameterize second vector as in the consideration above
$$
{m\over 2T}{r^2-{l^2\over M^2} \over (1-{m^2T^2\over \nu^2}+{{m\alpha\over r}-E_1)\over M^2}}
={1\over 2}{(s_0^1-(s_+^1e^{-a}+s_-^1e^a)\over \sinh a} , 
$$
$$
{T\over 2m}(-E_1+2m\epsilon+{m\alpha\over r}=s_+^1-s_-^1
$$
$$
{\dot {r^2}\over 1-{m^2T^2\over \nu^2}+{{m\alpha\over r}-E_1\over M^2}}
={[s_+^1-\cosh a s^1_0+s_-^1)]\over \sinh a}
$$
 
Three equations above is the linear system with respect to unknown components of the first vector $\rig{s^1}{}$. Matrix of this system
$$
M=\pmatrix{-{e^{-a}\over 2\sinh a} & {1\over 2\sinh a} & -{e^a\over 2\sinh a} \cr 
                                 1 & 0 & -1 \cr
-{1\over \sinh a} & {\cosh\over \sinh a} & -{1\over \sinh a} \cr}
$$
have the following inverse one
$$
M^{-1}=\pmatrix{{\cosh a\over \sinh a} & {1\over 2\sinh^2 a}-{e^a\cosh a\over 2\sinh^2 a} & -{1\over 2\sinh a} \cr 
              {2\over \sinh a} & -{1\over \sinh a}& -{\cosh\over \sinh a}\cr
                      {\cosh\over \sinh a} & -{1\over 2\sinh^2 a}+{e^{-a}\cosh\over 2\sinh^2 a} & -{1\over 2\sinh a} \cr}
$$
We remind that $ s_+^1s_-^1-{1\over 4}(s_0^1)^2={1\over 4}l^2$ and thus we have
$$
W^T(M^{-1})^T\pmatrix{ 0 & 0 & 1 \cr
                      0 & -{1\over 2} & 0 \cr
                      1 & 0 & 0 \cr}M^{-1}W={1\over 2}l^2
$$
where $W$ three dimensional vector of right-hand side of linear system, each component of which a
is expressed via only $r$ and its derivative of the first order. Calculations the matrix $Q$ of quadratical form leads to the result
$$
         Q=\pmatrix{ 2 & -1 & 0 \cr
                      -1 & 0 & 0 \cr
                      0 & 0 & -{1\over 2} \cr}
$$
and finally 
$$
-2[{m\over 2T}{r^2-{l^2\over M^2} \over (1-{m^2T^2\over \nu^2}+{{m\alpha\over r}-E_1)\over M^2}}]^2-{1\over 2}([{r^2-{l^2\over M^2} \over (1-{m^2T^2\over \nu^2}+{{m\alpha\over r}-E_1)\over M^2}}](E_1-2m\epsilon-{m\alpha\over r})
$$
\begin{equation}
-{1\over 2}[{\dot {r^2}\over 4(1-{m^2T^2\over \nu^2}+{{m\alpha\over r}-E_1)\over M^2}}]^2={1\over 2}l^2\label{BUIT}
\end{equation}
This equation of the first order with respect to $r$ exactly of same kind as equation of the Kepler problem in usual theory
$$
(\dot r)^2={2\epsilon \over m}-{2\alpha\over m r}-{l^2\over m^2r^2}
$$
We demonstrate the connection of the technique of the present paper with usual consideration in Appendix.

It is necessary keep in mind that in quantum space $t$ is also dynamical variable as coordinates and after integration in quadratures equation (\ref{BUIT}) it is necessary to find solution of equation (\ref{TR}) which connected proper time with the real time of quantum space. Indeed from (\ref{PR}) we obtain connection between $t,r$ and $a$ variables in a form
$$
-{M^2t^2\over 2T m}=s_-^2={l\over 2}l R e^{-a},\quad ({t-{m T^2\over \nu}\over T})^2=
-(1-{m^2T^2\over \nu^2}-{E_1-{m\alpha\over r}\over M^2}){e^{-a}\over 2\sinh a}
$$
Thus equation (\ref{BUIT}) is integrable in quadratures. The same is true with respect to equation for $a$ (\ref{TR}) and form the last equation we obtain conection between proper time and real time of the world.

\section{Outlook}

In the present paper we have solved Kepler problem in quantum non relativistic space,
equation of which are invariant with respect to Galilei group of motion. Thus this theory it is possible to consider as alternative to usual dynamics, as its generalization satisfying to general principles of Newton mechanics. We don't want to do some speculations and give any prediction before detail analyses which must be done for experimental testing of the theory proposed above. We want only pay attention of the reader that in resolving of the Kepler problem in deformed space we have two integrals of energy usual $\epsilon$ (as time shift in Galilei algebra) and ${E_1\over m}$, which arises in a process of integration and disapiere under limiting procedure passing to Newton space-time.

\section{Aknowledgements}

The author thanks CONACYT for finansical support.

\section{Appendix}

Kepler problem in Newton case is described by Hamiltonian function
$$
H={p^2\over 2m}+{\alpha\over r},\quad p^2=2mH-{2m\alpha\over r} 
$$
We have 
$$
\dot r^2=2r\dot r=[H,r^2]=[{p^2\over 2m},r^2]=2{(pr)\over m}
$$
$$
r^2(2mH-{2m\alpha\over r})-(mr\dot r)^2=p^2r^2-(pr)^2=l^2
$$
which is the main equation of the Kepler problem in usual theory.


\begin{thebibliography}{9}

\bibitem{0}A.N. Leznov  {\it Free motion in deformed (quantum) four-dimensional space time e-Print: hep-th/07073420}
\bibitem{SCH} Schrodinger E  {\it Proc.R.Irish.Acad. A 46, 9 (1940)}
\bibitem{IS} Infeld L. and Shild A. {\it Phys.Rev. 67, 121 (1945)} 
\bibitem{HIG}Higgs P.W.{\it J.Phys. A: Math.Gen 12,309 (1979)}
\bibitem{GKO} V.V.Gritzev, Yu.A.Kurochkin and V.S.Otchik
{\it J.Phys.A| Math.Gen.33(2000)4903-4910}.
\bibitem{I} Snyder {\it Phys.Rev. 71 (1947) 38}, Yang {\it Phys.Rev. 72 (1947) 874},
A.N.Leznov and V.V.Khrushov {\it Preprint INEP 73-99 (1973), Grav.Cosmol.9:159,2003.
e-Print: hep-th/0207082}.
\bibitem{LEZM} A.N. Leznov {\it Quantized spaces are four-dimensional compact manifolds with de-Sitter (O(1,4) or O(2,3)) group of motion.e-Print: hep-th/0410255 }
{\it  Theory of fields in quantized spaces. e-Print: hep-th/0409102}
\bibitem{LEZQ} A.N. Leznov {\it Nuclear Physics B640 [PM] (2002) 469-480}
\bibitem{LM} A.N. Leznov and J.Mostovoy {\it Classical dynamics in deformed spaces. e-Print: hep-th/0208152 Published in J.Phys.A36:1439-1450,2003}.
\end{thebibliography}
\end{document}